\documentclass[amsmath,amssymb,pra,twocolumn,longbibliography]{revtex4-2}
\usepackage{graphicx,amsmath,amsfonts,bm, color,MnSymbol}

\usepackage{bm}
\usepackage{hyperref}
\newcommand*\colvec[3][]{\begin{pmatrix}\ifx\relax#1\relax\else#1\\\fi#2\\#3\end{pmatrix}}
\newcommand*\rowvec[3][]{\begin{pmatrix}\ifx\relax#1\relax\else#1&\fi#2&#3\end{pmatrix}}

\newcommand{\Grad}{\nabla}

\newcommand{\Div}{\nabla\cdot}

\newcommand{\mycomment}[1]{}

\newcommand{\mcI}{\mathcal{I}}

\newcommand{\UCD}[1]
{\stackrel{\smalltriangledown}{#1}}

\begin{document}

\title{Polymer-polymer interdiffusion: \\ effects of entanglements and a polymeric source}
\author{Avraham Moriel and Howard A. Stone}
\affiliation{Department of Mechanical and Aerospace Engineering, Princeton University, Princeton, New Jersey 08544, USA}

\begin{abstract}
Many industrial applications and biological scenarios involve the interdiffusion of two polymeric species. Motivated by biological subcellular source-driven processes, we study polymer-polymer interdiffusion problems in the absence or the presence of a polymeric source, for both unentangled and entangled scenarios. Utilizing a two-fluid formalism, we arrive at scaling relations, self-similar reductions, and analytical solutions, which are confirmed with one- and two-dimensional numerical simulations. The introduction of a source term breaks the self-similar structure, modifying the boundary conditions and the domain of integration. Nevertheless, we show that the front characteristics of the diffusing droplet exhibit similar spatial structures as in the absence of a source. Our results allow deeper understanding of polymer-polymer interdiffusion and nonlinear transport, especially in the presence of a source.
\end{abstract}
\maketitle

\section{Introduction}\label{se:intro}

Polymers play a crucial role in many technological and biological applications~\cite{de1979scaling,rubinstein2003polymer,doi1988theory}. A single polymer exhibits complex dynamics that span multiple time and length scales~\cite{de1980dynamics,pincus1981dynamics,doi1978dynamics1,doi1978dynamics2,doi1978dynamics3,brochard1983polymer,brochard1991kinetics,brochard1983polymer,brochard1991kinetics,green1987diffusion,jordan1988mutual,brochard1986polymer,kausch1989polymer,klein1990interdiffusion,de1971reptation}. Multiple polymers put together can give rise to entanglements, leading to dynamics such as reptation~\cite{de1971reptation,klein1978evidence,russell1993direct}. Mixing two different polymeric species considerably complicates the dynamics, which subsequently depend on the underlying interactions between the polymer types, their degree of entanglements, as well as their mechanical properties. In such a problem, the polymers can phase-separate, or diffuse into one another --- a scenario termed polymer-polymer interdiffusion~\cite{de1980dynamics,pincus1981dynamics,doi1978dynamics1,doi1978dynamics2,doi1978dynamics3,brochard1983polymer,brochard1991kinetics,brochard1983polymer,brochard1991kinetics,green1987diffusion,jordan1988mutual,brochard1986polymer,kausch1989polymer,klein1990interdiffusion}. Below, we study the effects of entanglement on the spatio-temporal evolution of polymeric droplets undergoing interdiffusion.

Besides their technological value, concepts from polymer physics can help in understanding many biological systems. One of the predominant current frameworks to understand the formation of biological condensates is via liquid-liquid phase separation, which utilizes a similar framework as for polymer phase separation~\cite{brangwynne2015polymer,hyman2014liquid,brangwynne2009germline,shin2017liquid,berry2018physical}. For example, phase separation in polymer-fluid mixtures exhibits biologically relevant morphologies, such as gel networks and condensates~\cite{tanaka1993unusual,taniguchi1996network,tanaka1997phase,araki2001three,zhang2001kinetics,tanaka2006viscoelastic,doi2009gel,tanaka2022viscoelastic}. Consequently, the application of ideas from polymer science to biological scenarios has motivated measurements of mechanical properties of biological condensates, improving our understanding of their rheological responses~\cite{alshareedah2021programmable,jawerth2020protein,cheng2025micropipette,ghosh2021shear,ghosh2021shear}.

While passive processes such as diffusion and phase separation are ubiquitous, many biological processes involve active production of polymers, for example in transcription and translation. One prominent example of such an active source is the nucleolus --- a subnuclear condensate synthesizing and exporting ribosomal RNA (rRNA)~\cite{boisvert2007multifunctional,carmo2000or,riback2023viscoelasticity,mougey1993terminal,miller1981nucleolus}. The nucleolus acts as a persistent source, driving an outward flux of biopolymers into the surrounding nucleoplasm. Clearly, once such a source is present, the polymer concentration (of rRNAs in the above example) is also affected by the presence of the source.

\begin{figure*}[ht!]
\includegraphics[width=0.9\textwidth]{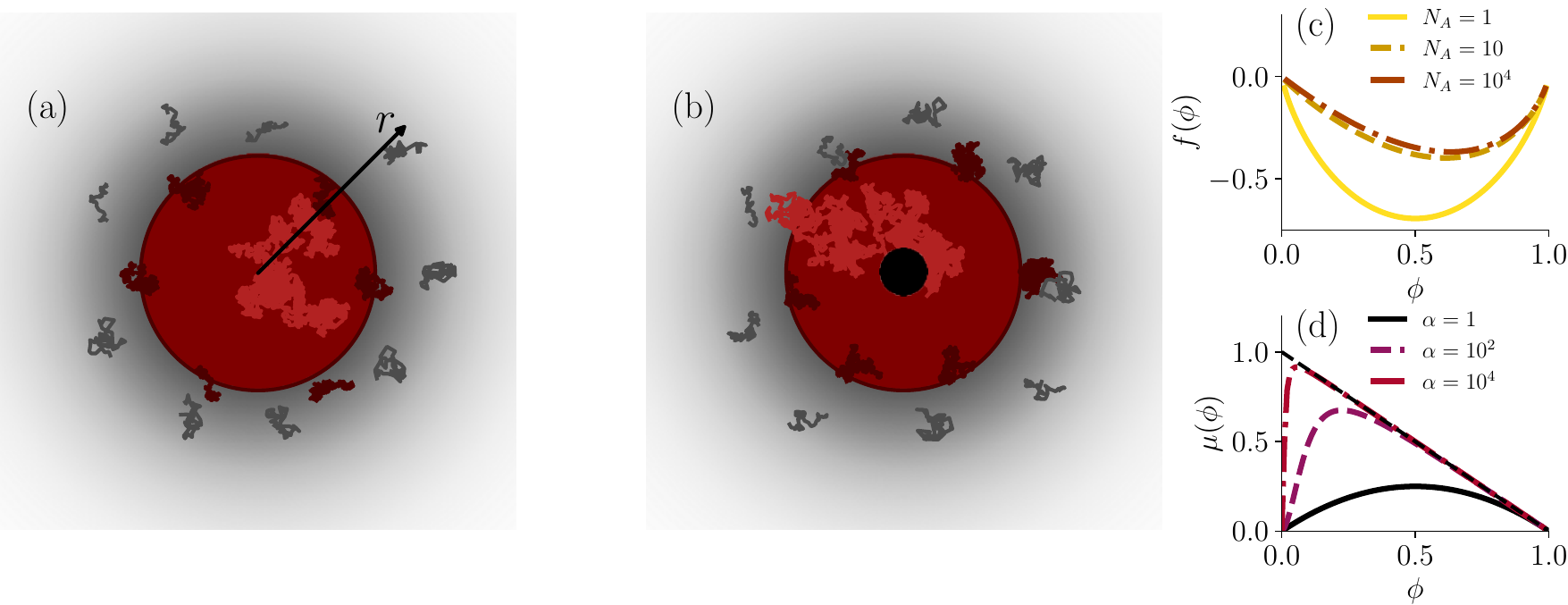}
\caption{Passive (droplet) diffusion, and source-driven diffusion. (a) A sketch of an A-mer droplet interdiffusing into a B-mer surrounding medium. The A-mers are depicted as red polymers, and B-mers as gray ones. (b) A source-driven problem, where the source is depicted as a small black circle at the center of the droplet. The source generates A-mers, that then diffuse into the B-mers. (c) Plots of $f(\phi)$ in the absence of interactions ($\chi\!=\!0$) and for monomer surroundings ($N_B\!=\!1$) for a wide range of polymer lengths $N_A$. Increasing the polymer length increases the energetic minima, and develops an asymmetry from $\phi\!=\!0.5$. (d) Plots of the friction function $\mu(\phi)$ for various values of $\alpha$ (the B-mers are assumed to be unentangled, $\mu_B\!=\!1-\phi$). As $\alpha$ increases, the friction function approaches $1-\phi$ (drawn as dashed black line).}
\label{fig:fig1}
\end{figure*}

Here, inspired by such biological condensates, we use a two-fluid formalism~\cite{milner1991hydrodynamics,doi1992dynamic,mavrantzas1992modeling,larson1992flow,beris1994thermodynamics,beris1994compatibility,tanaka1996universality,tanaka1997viscoelastic,tanaka2000viscoelastic,tanaka2012viscoelastic} to study two types of polymer-polymer interdiffusion problems --- one of passive droplet diffusion, and another of source-driven diffusion, both depicted schematically in Fig.~\ref{fig:fig1}(a)-(b) respectively. We first review in Sec.~\ref{se:eom} the equations of motion of the two-fluid formalism~\cite{milner1991hydrodynamics,doi1992dynamic,mavrantzas1992modeling,larson1992flow,beris1994thermodynamics,beris1994compatibility,tanaka1996universality,tanaka1997viscoelastic,tanaka2000viscoelastic,tanaka2012viscoelastic}, which yield a reduced set of one-dimensional (1D) equations of motion. Then we analyze the passive droplet diffusion cases in Sec.~\ref{se:droplet}. We show that under certain approximations, the polymer droplet interdiffusion is equivalent to a nonlinear diffusion problem, allowing us to cast the equations of motion in a self-similar form. We use the 1D formalism to obtain full numerical solutions, and compare those with the self-similar approach. We also demonstrate that our results hold in a two-dimensional (2D) case. 

Then, we introduce a source and analyze the emerging scaling, the resulting reduced equations of motion, and the corresponding dynamics in Sec.~\ref{se:source}. Specifically, we demonstrate that the source introduces an additional flux term, and breaks the self-similar structure of the passive problem. Consequently, instead of using a self-similar ansatz, we perform a perturbation around the expanding droplet's tip, and show that the local geometry at the front of a source-driven droplet resembles the one observed in the passive diffusion scenario. We compare our analytical predictions against the 1D and 2D full numerical solutions, demonstrating the spatial similarity to the passive case. Finally, we discuss our results and potential implications in Sec.~\ref{se:discussion}.

\section{Equations of motion}\label{se:eom}

We consider a two-component mixture, of polymeric species A and B, referred to as A-mers and B-mers, both of the same density~\cite{milner1991hydrodynamics,doi1992dynamic,mavrantzas1992modeling,larson1992flow,beris1994thermodynamics,beris1994compatibility,tanaka1996universality,tanaka2000viscoelastic,tanaka2012viscoelastic,tanaka1997viscoelastic}. With the A-mer and B-mer velocity fields $\bm{u}_A(\bm{r},t)$ and $\bm{u}_B(\bm{r},t)$, respectively 
[below we suppress the explicit $(\bm{r},t)$ dependence for readability], the volume fractions $\phi_A$ and $\phi_B$ evolve via the continuity equations
\begin{equation}\label{eq:con_clean}
\begin{aligned}
    \partial_t \phi_A + \nabla\cdot\left(\bm{u}_A \phi_A\right) &= S \ , \\
    \partial_t \phi_B + \nabla\cdot\left(\bm{u}_B \phi_B\right) &= 0,
\end{aligned}
\end{equation}
where here we introduced an A-mer source $S\!\equiv\!S\left(\bm{r},t\right)$. Because we consider a binary mixture, $\phi_A + \phi_B\!=\!1$, and we denote $\phi_A\!=\!\phi$.

The quasi-static momentum conservation balances the pressure $p$, the osmotic pressure $\bm{\Pi}$, and the material stresses $\bm{\sigma}_{A}$ and $\bm{\sigma}_{B}$ as~\cite{tanaka2000viscoelastic,tanaka2012viscoelastic}
\begin{equation}
    {\bm 0} = \nabla\cdot\left(-p\bm{I}-\bm{\Pi} + \boldsymbol{\sigma}_{A} + \boldsymbol{\sigma}_{B}\right)  \ , \label{eq:qs_clean}
\end{equation}
where $\bm{I}$ is the identity tensor.

The motion of one component immediately implies the relative motion of the other component. This relative motion is obtained as~\cite{doi1992dynamic,tanaka2000viscoelastic,beris1994thermodynamics}
\begin{equation}\label{eq:rel_clean}
    \bm{u}_A - \bm{u}_B = -\frac{1}{ n_0\zeta_0\mu}\left[\left(1-\phi\right)\nabla\cdot\left(\bm{\Pi}-\bm{\sigma}_{A}\right)+\phi\nabla\cdot\bm{\sigma}_{B}\right] \ ,
\end{equation}
where $n_0$ is the mixture's number density (monomers per volume), $\zeta_0$ is a monomer friction coefficient (assumed to be the same for both A-mers and B-mers), and $\mu\!\equiv\!\mu\left(\phi\right)$ is a dimensionless friction function that depends on the polymers and their surroundings~\cite{brochard1983polymer,brochard1991kinetics,doi1992dynamic}. The functional form of $\mu$ is expressed via the friction function of the A-mers and B-mers, $\mu_A$ and $\mu_B$ respectively, as 
\begin{equation}\label{eq:mu}
    \mu \!=\! \frac{\mu_A \mu_B}{\mu_A + \mu_B} \ .
\end{equation}
The individual friction functions $\mu_A$ and $\mu_B$ depend crucially on the polymer's lengths, $N_A$ and $N_B$, compared to an entanglement threshold $N_e$. Polymers shorter than $N_e$ are considered unentangled, implying $\mu_{(i)}\!=\!\phi_{(i)}$ (for species $i$). Entangled polymers are subjected to increased friction, $\mu_{(i)}\!=\!\alpha_{(i)}\phi_{(i)}$, where $\alpha_{(i)}\!\equiv\!N_{(i)}/N_e\!>\!1$~\cite{brochard1983polymer,brochard1991kinetics}.

The osmotic pressure $\bm{\Pi}$ originates from the polymer concentration, i.e., chemical potential, gradients. We utilize the Flory-Huggins (FH) thermodynamic potential~\cite{flory1953principles,huggins1942theory} $F_{FH}\!=\!k_B T n_0 f(\phi)$, where $k_B$ is the Boltzmann constant, $T$ is the temperature, and
\begin{equation}\label{eq:FH}
    f(\phi) = \frac{\phi}{N_A} \log \phi + \frac{\left(1-\phi\right)}{N_B} \log\left(1-\phi\right) + \chi \phi \left(1-\phi\right) \ .
\end{equation}
Here, $N_A$ and $N_B$ are the number of monomers per A-mer and B-mer respectively, and $\chi$ is an energy interaction parameter between the two species. As we are interested in the mixing of the two species, we omit surface tension. For simplicity, we consider non-interacting polymers, setting $\chi\!=\!0$. For the FH potential of Eq.~\eqref{eq:FH}, the resulting osmotic pressure is isotropic, obtained as $\bm{\Pi}\!=\!k_B T n_0\bm{I} \left(\phi\partial_\phi f-f\right)$~\cite{de1979scaling,doi1992dynamic}. Note that the osmotic pressure divergence $\Div\bm{\Pi}$ is equivalent to the spatial gradients of the chemical potential, as $\!\Grad\left[\phi\partial_\phi f-f\right]\!=\phi \Grad \left(\partial_{\phi} f\right)$ (we use $\bm{\Pi}$ to be consistent with the existing literature). We plot $f(\phi)$ for different values of $N_A$ (in the non-interacting limit $\chi\!=\!0$, and for monomer surroundings, $N_B\!=\!1$) in Fig.~\ref{fig:fig1}(c).

The stresses $\bm{\sigma}_{A}$ and $\bm{\sigma}_{B}$ evolve via an upper-convected Maxwell (UCM) model~\cite{tanaka2000viscoelastic,boyko2024perspective}, 
\begin{equation} \label{eq:UCM}
    \UCD{\bm{\sigma}}_{(i)} = G_{(i)}\left(\bm{L}_{(i)} + \bm{L}_{(i)}^T\right) - \frac{1}{\lambda_{(i)}}\bm{\sigma}_{(i)} \ ,
\end{equation}
where $i$ denotes either A or B (no summation implied), $\UCD{\bullet}\equiv\partial_t \bullet + \bm{u}_{(i)} \cdot \nabla \bullet -\left[\bm{L}_{(i)}^T\cdot \bullet + \bullet \cdot \bm{L}_{(i)}\right]$ is the upper-convected time derivative, $G_{(i)}$ is the shear modulus, $\lambda_{(i)}$ is the relaxation time, and $\bm{L}_{(i)}\!\equiv\!\nabla \bm{u}_{(i)}$ is the velocity gradient tensor for the $i$th component. While below we focus on a regime where the material stresses $\bm{\sigma}_A$ and $\bm{\sigma}_B$ could be safely neglected, we retain those for completeness.

To ease the analysis of the above equations of motion, we rescale Eqs.~\eqref{eq:con_clean}-\eqref{eq:UCM} with $\lambda_A$ as a time scale, $\sqrt{\frac{k_B T \lambda_A}{\zeta_0}}$ as a length scale, and $n_0 k_B T$ as an energy density scale. Thus, we arrive at a  unitless, 1-dimensional (1D) reduced form of the above equations, cast as ($i=A,B$) 
\begin{gather}
    \partial_t \phi + \partial_x\left(u_A \phi\right) = S \ , \label{eq:1d_conA}\\
    -\partial_t \phi + \partial_x\left[u_B \left(1-\phi\right)\right] = 0 \ , \label{eq:1d_conB} \\
    \partial_x\left(-p-\Pi + \sigma_{A} + \sigma_{B}\right)  = 0  \ , \label{eq:1d_mombal}\\
    u_A - u_B = -\frac{1}{ \mu}\left[\left(1-\phi\right)\partial_x\left(\Pi-\sigma_{A}\right)+\phi\partial_x\sigma_{B}\right] \ , \label{eq:1d_rel}\\
    \partial_t \sigma_{(i)} + u_{(i)} \partial_x \sigma_{(i)} = 2\left(\sigma_{(i)}+G\right)\partial_{x}u_{(i)}- \frac{1}{\lambda_{(i)}}\sigma_{(i)} \label{eq:1d_UCM}\ . 
\end{gather}

Below we study the passive diffusive dynamics, and source-driven diffusive dynamics, of asymmetric polymer mixtures, where we assume $N_A\!\gg\!N_B$. We also assume that $N_B$ is less than the entanglement threshold $N_e$, implying the B-mers are not entangled throughout the dynamics. To allow the A-mers to transition from a possibly entangled state (if they are sufficiently long) to an unentangled state, we use $\mu_A\!=\!\phi\left[\alpha \phi + \left(1-\phi\right)\right]$ as the A-mers friction function. This function reduces to $\mu_A\!=\!\phi$ in the unentangled limit (when $\alpha\!=\!1$). If the polymers are sufficiently long to entangle, it allows for a transition $\mu_A\!\rightarrow\!\phi$ at low volume fractions, and $\mu_A\!\rightarrow\!\alpha$ as $\phi\!\rightarrow\!1$. We plot three examples of $\mu(\phi)$ for different $\alpha$ values in Fig.~\ref{fig:fig1}(d). As $\alpha$ increases, $\mu$ develops an asymmetry, increasing the friction at intermediate $\phi$ regions and approaches $1-\phi$ --- an effect that is due to entanglements.

As we would like to focus on the role of entanglement and the passive or source-driven cases, we will neglect the roles of the material stresses $\bm{\sigma}_A$ and $\bm{\sigma}_B$. To estimate their relative importance, we note that polymeric stresses $\bm{\sigma}_{(i)}$ are usually characterized by the elastic moduli $G_{(i)}$, and relaxation times $\lambda_{(i)}$, as explained above~\cite{boyko2024perspective,tanaka2000viscoelastic,beris1994thermodynamics} [see also Eq.\eqref{eq:UCM}]. While precise measurements are challenging, biological condensates yield maximal storage moduli of $\sim\!1\!-\!10 \ \text{Pa}$ at the low-frequency limit, and maximal viscosity of $\sim\!10\!-\!100 \ \text{Pa} \cdot \text{s}$~\cite{alshareedah2021programmable,jawerth2020protein,cheng2025micropipette,ghosh2021shear,ghosh2021shear}. Comparing these estimates with the energy density scale $n_0 k_B T \simeq \text{KPa}$ at $T\!=\!300~K$ (using $n_0\!=\!1 ~\text{mM}$ as a low estimate for the monomer density), clearly indicates that under typical conditions the osmotic pressure is the dominant source of stress. 

The relaxation times of biopolymers may also be sufficiently fast for our interests. Biological polymers exhibit relaxation times of $\lambda\!\simeq\!10 \ \text{ms}$~\cite{alshareedah2021programmable,ghosh2021shear}. If we compare this to the expected (linear) diffusion time $\tau_D\!\simeq\!R^2 \zeta_0 / k_B T$ where $R$ is a typical droplet radius $R\!\simeq\!5 \ \mu \text{m}$, and $\zeta_0\!\simeq\!6\pi\eta a$ with $\eta\!=\!1 \ \text{mPa}\cdot\text{s}$ and $a\!\simeq\!1 \ \text{nm}$ being the monomer (protein) typical size, we get $\tau_D\!\simeq\!100 \ \text{ms}$, again confirming that polymer relaxation occurs at shorter times than those of interest here. Interestingly, a reminiscent time-scale separation was observed in~\cite{ghosh2021shear} for trapped droplets composed of different biopolymers and probed with optical tweezers. These separations of stress scales and time scales allow us to confine our analysis to cases in which these two assumptions hold, and the material stresses $\bm{\sigma}_A$ and $\bm{\sigma}_B$ can be safely neglected. While velocity gradients and polymer conformations near interfaces could induce mechanical stresses, considering the osmotic pressure captures the dominant driving force needed to describe the emerging nonlinear transport. As we show below, the simplifying assumptions above allow us to study polymer-polymer interdiffusion and its dependence on two central aspects --- the degree of entanglement and the absence or presence of a source.


\section{Droplet diffusion}\label{se:droplet}
We first consider a droplet of A-mers placed in a B-mers medium, depicted schematically in Fig.~\ref{fig:fig1}(a). As the species are assumed to be non-interacting ($\chi\!=\!0$), entropic forces cause the two species to diffuse into one another. This process is characterized by an effective nonlinear diffusion equation.

To see this description, we utilize our 1D formalism, and add Eq.~\eqref{eq:1d_conA}-\eqref{eq:1d_conB}. We define $v\!\equiv\! u_A\phi+u_B(1-\phi)$, which can be interpreted as the mean velocity, and obtain $\partial_x v\!=\!0$ in the absence of a source ($S\!=\!0$), implying $v\!=\!0$ due to isotropy. We define the relative speed $w\!\equiv\!u_A- u_B$, and using $u_A\!=\!v+(1-\phi)w$, we obtain that in the absence of a source $u_A\!=\!(1-\phi)w$. We then use Eq.~\eqref{eq:1d_rel} with Eq.~\eqref{eq:1d_conA} (neglecting the material stresses $\sigma_A$ and $\sigma_B$), to write 
\begin{equation}\label{eq:droplet_1d}
    \partial_t \phi = \partial_x\left[\frac{\phi(1-\phi)^2}{\mu}\partial_x\Pi\right] \ , 
\end{equation}
which corresponds to a nonlinear diffusion equation, with
\begin{equation}\label{eq:osmotic_1d}
    \Pi\!=\!-\left[\log(1-\phi)+\left(1-N_A^{-1}\right)\phi\right] \ ,
\end{equation}
where we set $N_B\!=\!1$ and $\chi\!=\!0$, and 
\begin{equation}\label{eq:mu_exp}
 \mu = \phi\left(1-\phi\right)\frac{1-\phi+\alpha\phi}{1-\phi+\phi\left(1-\phi+\alpha\phi\right)}
\end{equation}
[see Eq.~\eqref{eq:mu}].



\begin{figure}[t]
\includegraphics[width=0.48\textwidth]{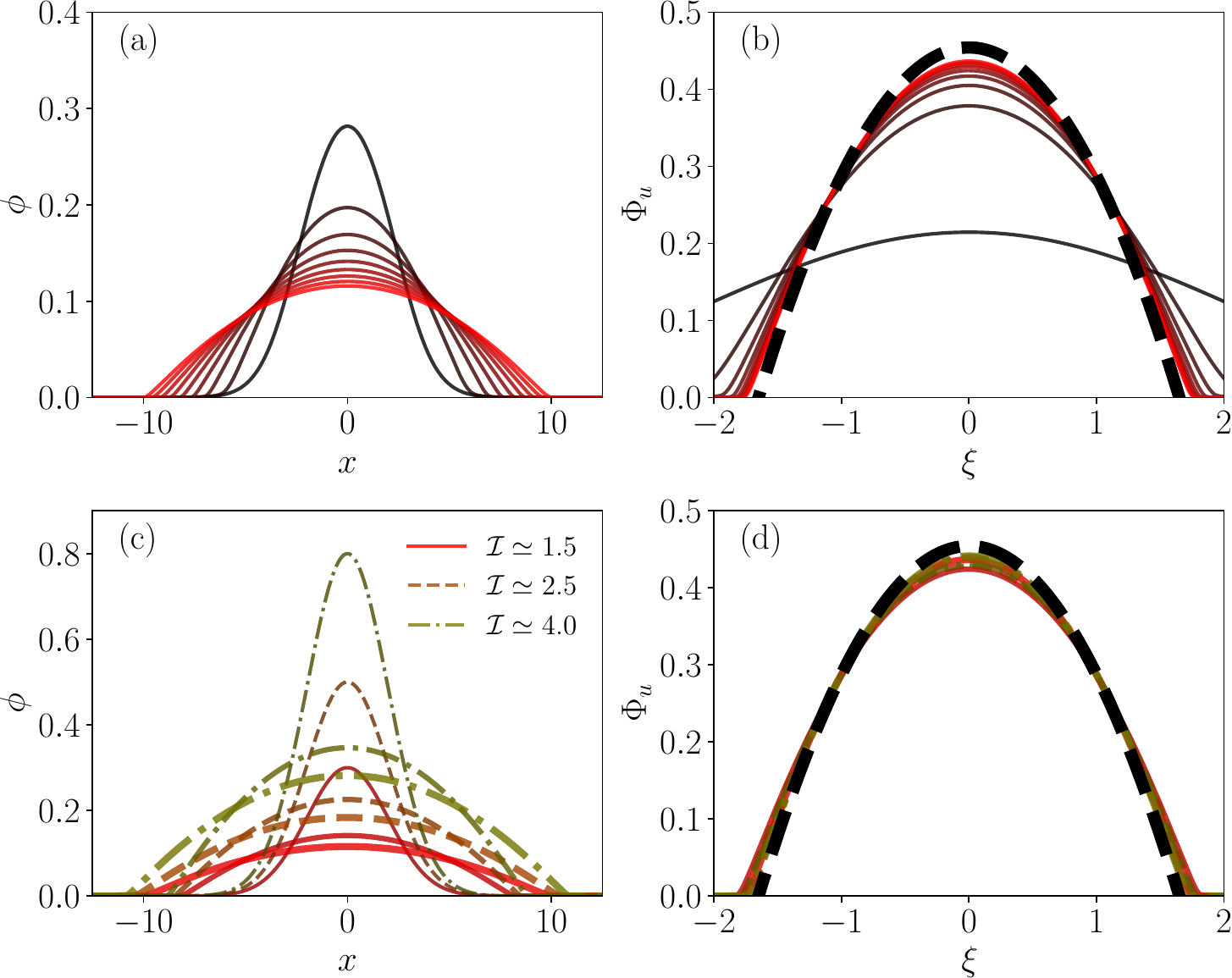}
\caption{Numerical and analytical solutions for unentangled polymer interdiffusion with $\alpha\!=\!1$. (a) The spreading A-mer droplet dynamics of $\phi$, as obtained from our 1D numerical simulations of Eqs.~\eqref{eq:1d_conA}-\eqref{eq:1d_UCM} $\phi_{\text{max}}\!=\!0.3$ corresponding to $\mcI\!\simeq\!1.5$). Early times are depicted as gray, and later times are drawn in red. (b) The same $\phi$ profiles  drawn using the self-similar $\Phi_u\!=\!\phi / \left(\mathcal{I}^2/t\right)^{1/3}$, and $\xi\!=\!x / \left(\mathcal{I}t\right)^{1/3}$. Dashed black line is the analytical solution $\Phi_u\!=\!-\frac{\xi^2}{6}+\left(\frac{3}{32}\right)^{1/3}$. (c) Snapshots of $\phi$ from simulations with different initial $\phi_\text{max}$ values of $0.3,\ 0.5$, and $0.8$, corresponding to different $\mathcal{I}$ values (see legend). (d) The same snapshots cast as self-similar solutions, with black dashed line the same as in (b).}
\label{fig:fig2}
\end{figure}

In the unentangled limit, we set $\alpha\!=\!1$, reducing $\mu_A\!=\!\phi$, and $\mu_B\!=\!1-\phi$, and from Eq.~\eqref{eq:mu_exp} $\mu\!=\!\phi(1-\phi)$ [see also Fig.~\ref{fig:fig1}(d) for a plot of $\mu(\phi)$ for different $\alpha$]. Using the 1D osmotic pressure of Eq.~\eqref{eq:osmotic_1d}, and $N_A\!\gg\!1$, Eq.~\eqref{eq:droplet_1d} becomes
\begin{equation}\label{eq:drop_1d_unentangled}
    \partial_t \phi \!\simeq \!\partial_x\left[\frac{(1 + N_A \phi)}{N_A}\partial_x\phi\right] \underset{\phi\gg N_A^{-1}}{\rightarrow} \partial_x\left(\phi \partial_x \phi\right)\ .
\end{equation}
For sufficiently low volume fractions $\phi N_A\!\ll\!1$, the above equation reduces to a linear diffusion equation. However, supposing that the initial volume fraction within the droplet is such that $\phi\!\gg\!N_A^{-1}$, Eq.~\eqref{eq:drop_1d_unentangled} approaches a nonlinear diffusion equation, as indicated above.

Using $\ell$ as a typical length scale, Eq.~\eqref{eq:drop_1d_unentangled} suggests $\ell^2\!\sim\!\phi t$. We combine this scaling relation with fact that the total volume fraction $\mathcal{I}\!\equiv\!\int_{-\ell}^{\ell}\phi dx$ must be conserved throughout the dynamics, implying $\ell \phi\!\sim\! \mathcal{I}$. Put together, these scaling relations suggest $\ell\!\sim\!\left(\mathcal{I}t\right)^{1/3}$, and $\phi\!\sim\!\left(\mathcal{I}^2/t\right)^{1/3}$.

We now use the self-similarity ansatz, $\phi(x,t)\!\equiv\!\left(\mathcal{I}^2/t\right)^{1/3}\Phi_u\left(\xi\right)$ (where $\Phi_u$ denotes the unentangled $\Phi$), with the self-similar coordinate $\xi\!\equiv\!x / \left(\mathcal{I}t\right)^{1/3}$. Substituting into the approximate nonlinear diffusion equation, Eq.~\eqref{eq:drop_1d_unentangled}, yields an ordinary differential equation (ODE),
\begin{equation}\label{eq:drop_1d_unentangled_ODE}
    \left(3\Phi_u\Phi_u'+\xi \Phi_u\right)'\!=\!0 \ .
\end{equation}
Integrating once, assuming a symmetric droplet profile $\Phi_u'(0)\!=\!0$, we arrive at $\Phi_u\!=\!-\frac{\xi^2}{6}+c$, where $c$ is an integration constant. The obtained solution is a parabola that intersects the $\xi$ axis $\Phi_u(\xi_*)\!=\!0$, at $\xi_*\!=\!\sqrt{6c}$. Integrating over the domain and demanding the volume fraction conservation $\int_0^{\xi_*}\Phi_u\!=\!\frac{1}{2}$ gives $c\!=\!\left(\frac{3}{32}\right)^{1/3}$.

We simulate the 1D system of Eqs.~\eqref{eq:1d_conA}-\eqref{eq:1d_UCM}, with a finite value of $N_A\!=\!10^4$ and setting $\alpha\!=\!1$ (using $N_B\!=\!1$, $\chi\!=\!0$ , and in the absence of $\sigma_A$ and $\sigma_B$). We initialize the distribution of $\phi$ as a Gaussian of typical length of $2$, and normalize it such that at its peak, $\phi\!=\!\phi_{\text{max}}$, i.e., $\phi(x,t=0)\!=\!\phi_{\text{max}}\exp(-x^2/8)$, and follow its diffusion. We then solve Eqs.~\eqref{eq:1d_conA}-\eqref{eq:1d_UCM}, first by calculating the velocity profiles $v$ and $w$, followed by integrating $\phi$ in time~\cite{van1995python,harris2020array}. 

We show in Fig.~\ref{fig:fig2}(a) the $\phi$ profiles obtained at different times during the diffusion dynamics. Then, we apply the self-similar transformation, and re-plot the solution in terms of the self-similar function $\Phi_u$, and variable $\xi$ in Fig.~\ref{fig:fig2}(b). We also plot the theoretical solution $\Phi_u$ on top. As time advances, the profiles approach the predicted solution, demonstrating the validity of the approximate nonlinear diffusion equation Eq.~\eqref{eq:drop_1d_unentangled} for finite $N_A$ values, and the emerging scaling relations above. In Fig.~\ref{fig:fig2}(c) we plot trajectories of $\phi$ profiles obtained for three different $\mathcal{I}$ values (modified by changing $\phi_{\text{max}}$). Rescaling those profiles as shown in Fig.~\ref{fig:fig2}(d), reveals the scalings and analytical solution applies to those different parameter values and at different times.


\begin{figure}[t]
\includegraphics[width=0.48\textwidth]{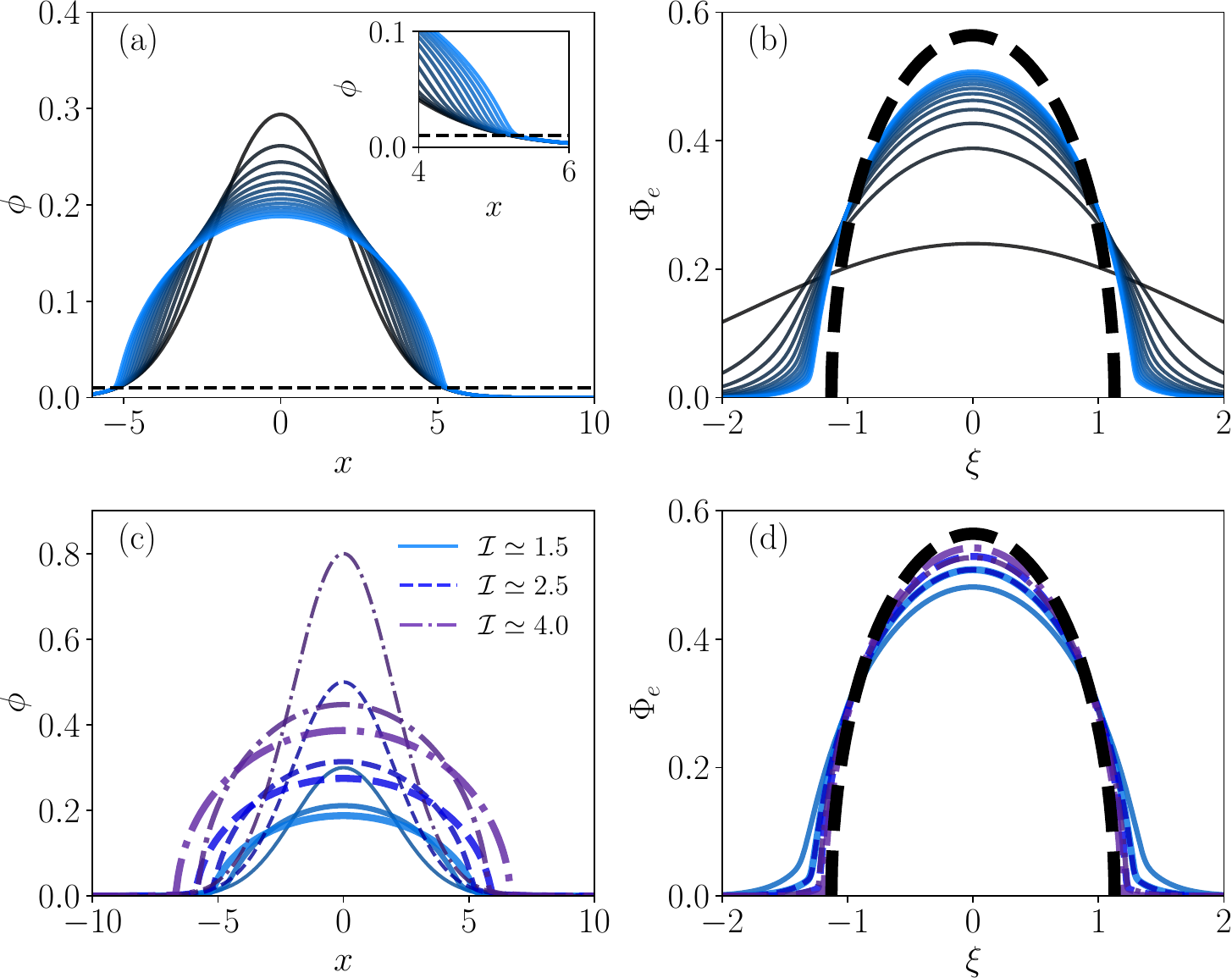}
\caption{Numerical and analytical solutions for entangled polymer interdiffusion with $\alpha\!=\!10^4$. (a) The spreading A-mer droplet dynamics of $\phi$, as obtained from our 1D numerical simulations of Eqs.~\eqref{eq:1d_conA}-\eqref{eq:1d_UCM} (using $\phi_{\text{max}}\!=\!0.3$ corresponding to $\mcI\!\simeq\!1.5$). Early times are depicted as gray, and later times are drawn in blue. Horizontal dashed line indicates $\phi\!=\!\alpha^{-1/2}$. Inset: a close-up on the transition region $\phi\!\simeq\!\alpha^{-1/2}$ showing the formed tails. (b) The same $\phi$ profiles  drawn using the self-similar $\Phi_e\!=\!\phi / \left(\mathcal{I}^2/t\right)^{1/4}$, and $\xi\!=\!x / \left(\mathcal{I}^2t\right)^{1/4}$. Dashed black line is the analytical solution $\Phi_e\!=\!\frac{1}{2}\sqrt{\frac{4}{\pi}-\xi^2}$. (c) Snapshots of $\phi$ from simulations with different initial $\phi_\text{max}$ values of $0.3,\ 0.5$ and $0.8$, corresponding to different $\mathcal{I}$ values (see legend). (d) The same snapshots cast as self-similar solutions, with black dashed line the same as in (b).}
\label{fig:fig3}
\end{figure}

We repeat a similar analysis for the case of entangled A-mers. For $\alpha\!\gg\!1$, we obtain $\mu\!\rightarrow\!1-\phi$ [see Fig.~\ref{fig:fig1}(d)], and Eq.~\eqref{eq:droplet_1d} reduces to
\begin{equation} \label{eq:drop_1d_entangled}
    \partial_t \phi \!\simeq \!\partial_x\left[\frac{(1 + N_A \phi)}{N_A}\phi \partial_x\phi\right] \underset{\phi\gg N_A^{-1}}{\rightarrow} \partial_x\left[ \phi^2 \partial_x \phi\right]\ .
\end{equation}
In this case, the scaling relations are $\ell^2\!\sim\!\phi^2 t$ and $\phi \ell \!=\! \mathcal{I}$, yielding $\ell\!\sim\!\left(\mathcal{I}^2t\right)^{1/4}$, and $\phi\!\sim\! \left(\mathcal{I}^2/t\right)^{1/4}$. Using the ansatz $\phi(x,t)\!\equiv\!\left(\mathcal{I}^2/t\right)^{1/4} \Phi_e\left(\xi\right) $ (here with a subscript $\Phi_e$ to denote the entangled case) with $\xi\!\equiv\!x / \left(\mathcal{I}^2t\right)^{1/4}$, we reduce Eq.~\eqref{eq:drop_1d_entangled} to
\begin{equation}\label{eq:drop_1d_entangled_ODE}
    \left(4\Phi_e^2\Phi_e'+\xi \Phi_e\right)'\!=\!0 \ .
\end{equation}

Using a similar procedure as described above we obtain $\Phi_e\!=\!\frac{1}{2}\sqrt{\frac{4}{\pi}-\xi^2}$. We show in Fig.~\ref{fig:fig3} numerical solutions for a 1D system in the entangled case in similar fashion to Fig.~\ref{fig:fig2} (with $N_A\!=10^4$ and $\alpha\!=\!10^4$). Specifically, Fig.~\ref{fig:fig3}(a) shows an exemplary trajectory at different times, and the inset shows a cusp occurring once $\phi\!\simeq\!\alpha^{-1/2}$ --- a ``tail'' where the simplification to $\phi^2\partial_x\phi$ is no longer valid, and a different scaling law emerges. This tail signifies a transition between an entangled state to the unentangled state, characterized by a different diffusion scaling. Then, Fig.~\ref{fig:fig3}(b) shows the collapse of the same trajectory using the suggested scalings, with the theoretical solution on top.  


The agreement between the full numerical solution and the self-similar solution is partial --- the droplet obtained from the numerical solutions of Eqs.~\eqref{eq:1d_conA}-\eqref{eq:1d_UCM} spreads further, and the ODE solution predicts higher central concentrations and lower concentrations at the rims. We suspect this partial agreement is due to the tails referred to above. As these tails move material away from the central droplet, we anticipate that as their relative contribution decreases, the solutions should approach the self-similar solution of Eq.~\eqref{eq:drop_1d_entangled_ODE}. In Fig.~\ref{fig:fig3}(c) we show solutions with different $\mathcal{I}$ (obtained by changing the $\phi_{\text{max}}$ values), and in Fig.~\ref{fig:fig3}(d) we show their rescaled counterparts. As the initial maximal concentration $\phi_{\text{max}}$ is increased, $\mcI$ increases, and the relative contribution of the tail to the dynamics decreases. The rescaled solutions indeed confirm that as the tail contributions become negligible (higher $\phi_{\text{max}}$ and $\mcI$ values) the solutions approach the expected self-similar form.

\begin{figure}[t]
\includegraphics[width=0.48\textwidth]{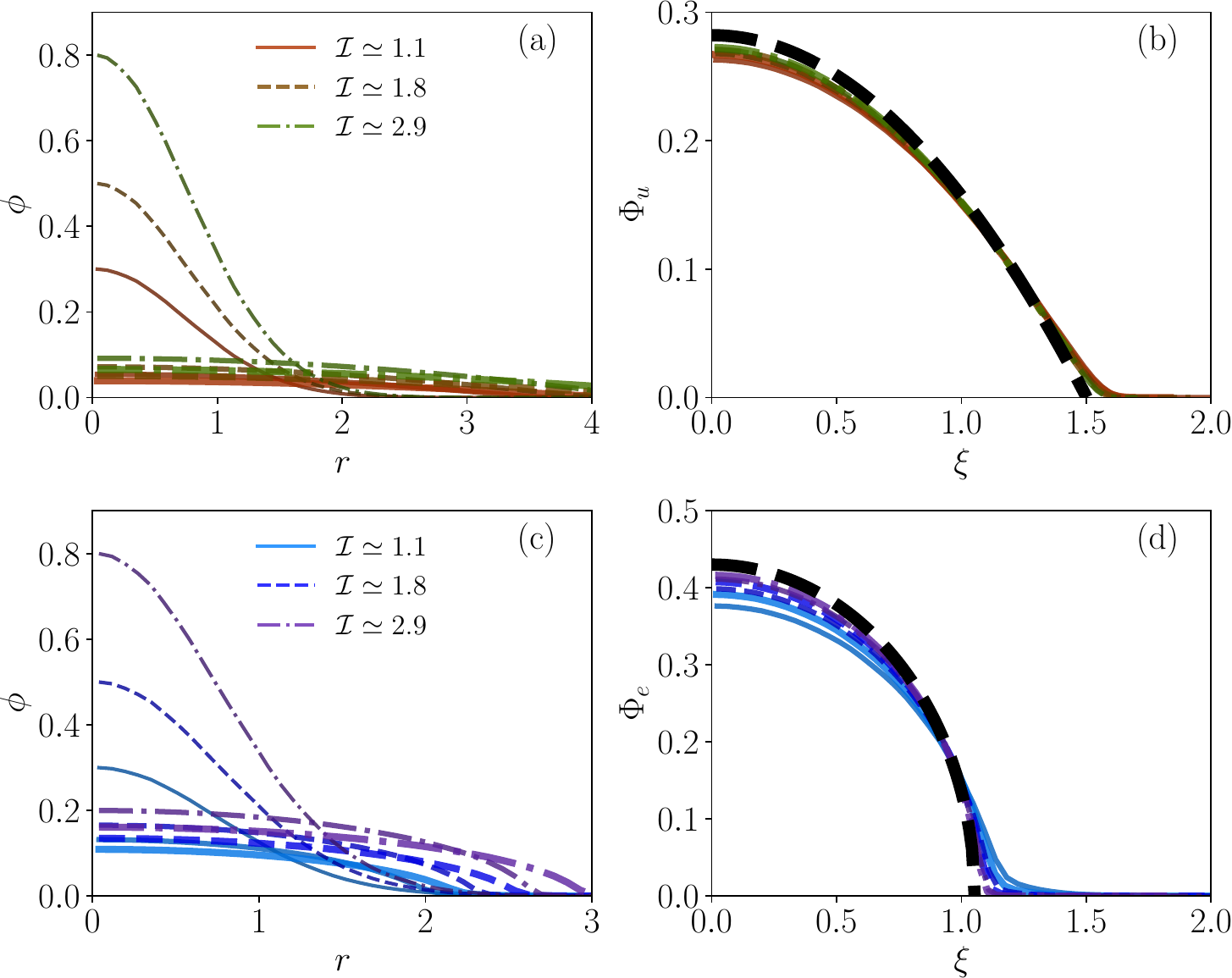}
\caption{Numerical and analytical solutions for unentangled polymer interdiffusion in 2D. (a) Snapshots of the radially-averaged $\phi$ from 2D simulations of Eqs.~\eqref{eq:con_clean}-\eqref{eq:rel_clean}, with $\alpha\!=\!1$, and with different initial $\phi_\text{max}$ values of $0.3,\ 0.5$ and $0.8$, corresponding to different $\mathcal{I}$ values (see legend). (b) The same snapshots as in (a) cast as self-similar solutions, with the self-similar solution plotted as a thick dashed black line. (c) Same as panel (a) but for the entangled case with $\alpha\!=\!10^4$. (d) The same snapshots as in (c) cast as self-similar solutions, with the self-similar solution in black dashed line.}
\label{fig:fig4}
\end{figure}

In two dimensions (2D) we can repeat the procedure above, to approximate the problem as a nonlinear diffusion equation, which could be cast in a self-similar framework (see App.~\ref{ap:2D_formalism}). Using $\bm{r}$ as the position vector, and assuming isotropy, we use the radial distance $r\!\equiv\!|\bm{r}|$ to get $\phi(r,t)\!\equiv\!\left(\mathcal{I}/t\right)^{1/2} \Phi_u(\xi)$, and $\xi\!\equiv\!r / \left(\mathcal{I}t\right)^{1/4}$, and $\phi (r,t)\!\equiv\!\left(\mathcal{I}/t\right)^{1/3} \Phi_e(\xi)$, and $\xi\!\equiv\!r / \left(\mathcal{I}^2 t\right)^{1/6}$ for the unentangled and entangled cases, respectively. Using the transformations we obtain
\begin{gather}
    \xi^2 \Phi_u + 4 \xi \Phi_u \Phi_u' = 0 \ \quad   \text{for} \ \alpha\!=\! 1 \ , \label{eq:2d_diffusion_unentangled} \\
    \xi^2 \Phi_e + 6 \xi \Phi_e^2 \Phi_e' = 0 \ \quad   \text{for} \ \alpha\!\gg\! 1 \ .\label{eq:2d_diffusion_entangled} 
\end{gather}
Solving those in a similar fashion to the 1D case yields $\Phi_u\!=\!\frac{1}{2}\left(\frac{1}{\sqrt{\pi}} - \frac{\xi^2}{4}\right)$, and $\Phi_e\!=\!\frac{1}{2}\sqrt{\left(\frac{2}{\pi}\right)^{2/3} - \frac{2\xi^2}{3}}$. We show our numerical solutions to Eqs.~\eqref{eq:con_clean}-\eqref{eq:rel_clean} (neglecting $\bm{\sigma}_A$ and $\bm{\sigma}_B$), and the comparison to the expected self-similar solutions in Fig.~\ref{fig:fig4}.

Analysing the passive diffusion problems, we note the similarity in the attained spatial form between the 1D and 2D cases, i.e., between Figs.~\ref{fig:fig2},~\ref{fig:fig3}, and ~\ref{fig:fig4}. Specifically, the unentangled cases vary with $\xi^2$, and the entangled cases vary as $\sqrt{\xi_*^2-\xi^2}$. Those similar spatial profiles hint that dimensionality plays a rather minor role in terms of the observed spatial profiles.

Generically, the suggested simplifications reduce both the 1D and 2D cases to a nonlinear diffusion problem with $\partial_t \phi \!=\!\nabla\cdot \left(\phi^{m-1}\nabla \phi\right)$, where $m$ quantifies the degree of nonlinearity that directly depends on $\alpha$ --- the unentangled case implies $\alpha\!=\!1$ leading to $m\!=\!2$, while the entangled case has $\alpha\!\gg\!1$ leading to $m\!=\!3$. Interestingly, similar equations appear in the context of diffusion in porous media~\cite{barenblatt2003scaling,zheng2022influence}. The diffusive self-similar length-scale and amplitudes depend on the nonlinearity $m$, and the dimensionality $d$ of the problem. A passive droplet diffusion problem in $d$ dimensions, with a nonlinearity of power $m$, suggests the scaling $\phi\!\sim\!\mathcal{I}^{\gamma_1}t^{\delta_1}$ and $\ell\!\sim\!\mathcal{I}^{\gamma_2}t^{\delta_2}$, with
\begin{equation}
\label{eq:scaling_diffusion}
\begin{aligned}
    &\gamma_1\!=\!2/p \ , &  \, \delta_1\!=\!-d/p \ , & \\
    &\gamma_2\!=\!(m-1)/p \ ,&  \, \delta_2\!=\!1/p \ ,& 
\end{aligned}
\end{equation}
with $p\!\equiv\!2+d(m-1)$. Transitioning from the unentangled to entangled scenario slows the dynamics down --- $m$ increases causing $p$ to increase as well. This parallels with polymer reptation in entangled scenarios, absent from the unentangled case~\cite{de1971reptation,brochard1983polymer,brochard1991kinetics}. So, while reptation is a single polymer mode-of-motion, it affects the emerging diffusion on a coarse-grained scale. In both cases, approximating the equations of motion in the limit of sufficiently large $N_A$ and $\phi$, and using the corresponding $\alpha$ values (while neglecting the material stresses $\sigma_A$ and $\sigma_B$) allowed us to reformulate the interdiffusion problem as a self-similar droplet diffusion problem. The emerging equations were simple enough to solve analytically, and resulted in good agreement with both our 1D and 2D numerical solutions.  

\section{Source-driven diffusion}\label{se:source}
The passive droplet diffusion scenario investigated above may become modified in certain biological cases. Some biological condensates serve as effective sources, e.g., in the synthesis of RNAs and proteins, injecting polymers at a certain rate into the system~\cite{boisvert2007multifunctional,carmo2000or,riback2023viscoelasticity,mougey1993terminal,miller1981nucleolus}. We now modify the formalism above to account for these scenarios, treating the synthesis centers as external polymer sources, depicted schematically in Fig.~\ref{fig:fig1}(b). 

\begin{figure}[t]
\includegraphics[width=0.48\textwidth]{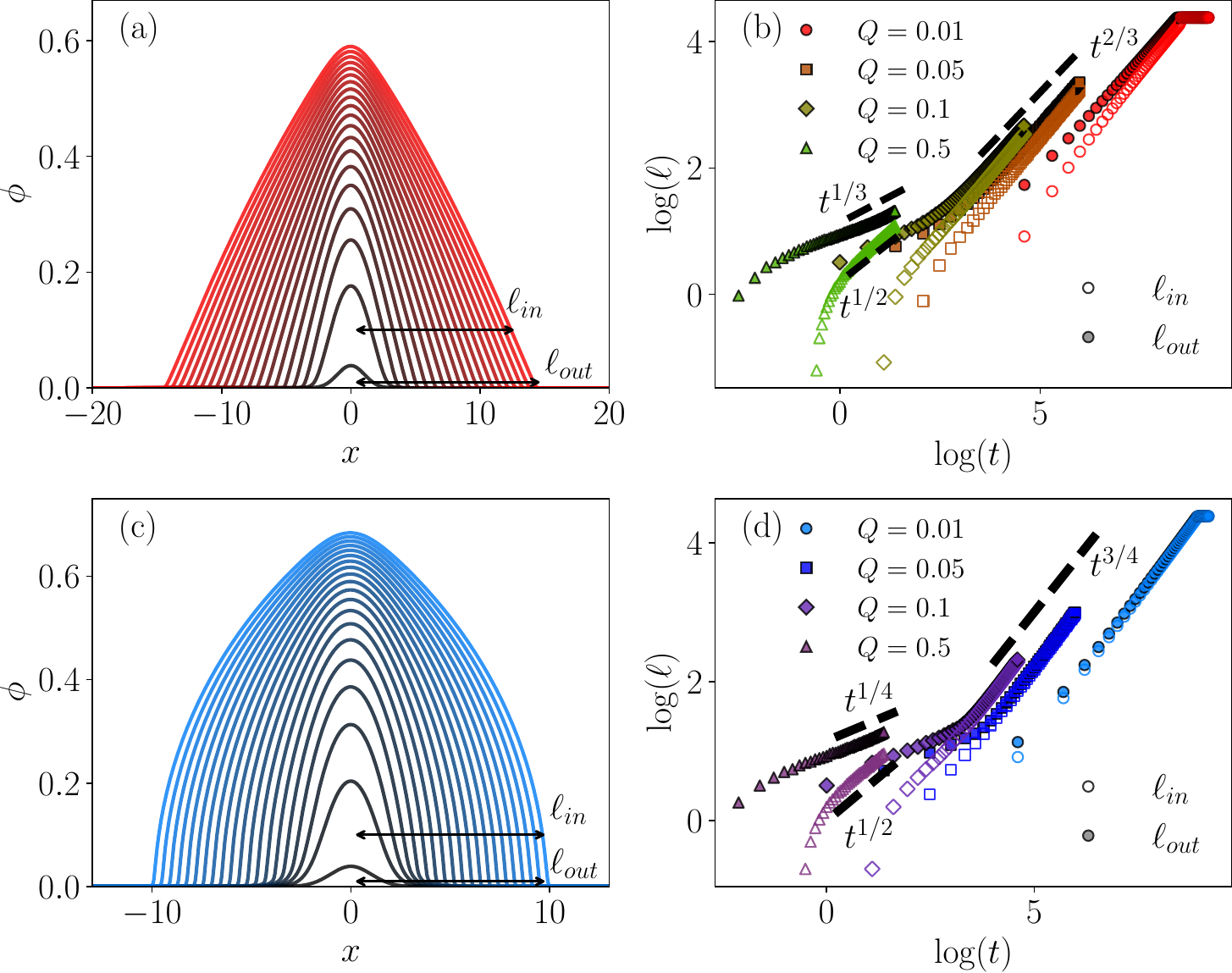}
\caption{Numerical solutions for unentangled and entangled source-driven polymer interdiffusion with $f(t)\!=\!Q$. (a) Unentangled $\alpha\!=\!1$ source-driven A-mer dynamics $\phi$, as obtained from our full 1D numerical simulations of Eqs.~\eqref{eq:1d_conA}-\!\eqref{eq:1d_UCM} (with $\alpha\!=\!1$, $N_A\!=\!10^4$, and $Q\!=\!0.1$). Early times are depicted as gray, and later times are drawn in red. We denote the two length scales $\ell_{in}$ and $\ell_{out}$, as the half-width of the droplet's domain that is $\phi\!>\!0.1$ for $\ell_{in}$ and $\phi\!>\!0.01$ for $\ell_{out}$. (b) Time-dependence of $\ell_{in}$ and $\ell_{out}$ for different source amplitudes $Q$. At short times, $\ell_{in}\!\sim\!t^{1/2}$ similar to a linear diffusion scaling,  while $\ell_{out}\!\sim\!t^{1/3}$ as in the passive droplet scenario. At long times, both length scales converge to a nonlinear scaling $\sim\!t^{2/3}$. (c) Entangled source-driven A-mer dynamics $\phi$, as obtained from our full 1D numerical simulations of Eqs.~\eqref{eq:1d_conA}-\!\eqref{eq:1d_UCM} (with  $\alpha\!=\!10^4$, $N_A\!=\!10^4$, and $Q\!=\!0.1$). Early times are depicted as gray, and later times are drawn in blue. We denote the two length scales $\ell_{in}$ and $\ell_{out}$ [following similar definitions as in panel (a)]. (d) Time-dependence of $\ell_{in}$ and $\ell_{out}$ for different source amplitudes $Q$. At short times, $\ell_{in}\!\sim\!t^{1/2}$ similar to a linear diffusion scaling,  while $\ell_{out}\!\sim\!t^{1/4}$ as in the passive droplet scenario. At long times, both length scales converge to a nonlinear scaling $\sim\!t^{3/4}$.}
\label{fig:fig5}
\end{figure}

We employ a similar numerical solution method as for the passive case discussed in Sec.~\ref{se:droplet} to obtain numerical solutions for the source-driven diffusion. Here, we set $\phi(x,t=0)\!=\!0$, and use a localized Gaussian as our source term, $S(x,t)\!=\!\frac{f(t)}{Z} \exp(-x^2/2)$, with the normalization factor $Z\!\equiv\!\sum\exp(-x^2/2)\ \Delta x$ over our integration domain (here $\Delta x$ is the resolution of the discretized domain). We show two exemplary trajectories for the unentangled $\alpha\!=\!1$ case in Fig.~\ref{fig:fig5}(a), and $\alpha\!\gg\!1$ in Fig.~\ref{fig:fig5}(c), for a constant-rate, localized source $f(t)\!=\!Q$, as obtained from our numerical simulations of Eqs.~\eqref{eq:1d_conA}-\!\eqref{eq:1d_UCM}. The A-mers accumulate near the origin, increasing $\phi$ locally, hence increasing the osmotic pressure/chemical gradients, driving outward diffusion. The volume fraction profile evolves from its early time behavior, until it converges to the long time profile. While there are intricate spatial features at early times, and near the origin, e.g., the presence of a ``tip'', in both cases the volume fraction distributions near the edges, or droplet's ``front shape", attained at long times are reminiscent of their spatial features in the passive droplet expansion (see Figs.~\ref{fig:fig2}-\ref{fig:fig3}). 

To characterize the volume fraction redistribution dynamics, we measure two typical droplet radii, denoted by $\ell_{in}$ and $\ell_{out}$, defined here, respectively, as the half-width of the $\phi\!>\!0.1$ domain for $\ell_{in}$, and $\phi\!>\!0.01$ domain for $\ell_{out}$ [as annotated in Fig.~\ref{fig:fig5}(a) and (c)]. We plot their time-dependencies for various source rates $Q$, as shown in Fig.~\ref{fig:fig5}(b) and (d) for the unentangled and entangled cases, respectively. At very early times $\ell_{in}$ approximately increases according to a linear diffusion equation response $\sim\!t^{1/2}$, while $\ell_{out}$ diffuses according to the passive droplet cases analyzed in Sec.~\ref{se:droplet}, i.e., following Eq.~\eqref{eq:scaling_diffusion}. At longer times, the two radii converge into similar nonlinear scalings. These early-time, spatially-dependent scalings indicate the volume fraction redistributes heterogeneously throughout the droplet.


To try and understand these spatial and temporal behaviors, we revisit the continuity equations Eq.~\eqref{eq:1d_conA}-\eqref{eq:1d_conB}. Upon addition, and using $v\!\equiv\!\phi u_A + (1-\phi) u_B$, we obtain $\partial_x v\!=\!S$. We then use this together with $w\!=\!u_A-u_B$, to express $u_A$, and substitute back into Eq.~\eqref{eq:1d_conA}, leading to the evolution equation 
\begin{equation}\label{eq:source_1d}
    \partial_t \phi = \partial_x\left[\frac{\phi\left(1-\phi\right)^2}{\mu}\partial_x \Pi - v \phi \right] + S \ ,
\end{equation}
Compared to the passive droplet equivalent, Eq.~\eqref{eq:droplet_1d}, the additional source term not only contributes by adding more A-mers into the system, but also modifies the average velocity $v$ and contributes a flux term. We expect a similar effect happens in higher dimensions as well, as a consequence of a material source near the origin (see App.~\ref{ap:higherDs}).

We now suppose that the source is spatially-localized, and takes the general form $S(x,t)\!=\!\delta(x) f(t)$. Combining the unentangled and entangled cases, in the limit of $\phi\!\gg\!N_A^{-1}$, Eq.~\eqref{eq:source_1d} reduces to
\begin{equation}\label{eq:source_1d_simplified}
\partial_t \phi + v \partial_x \phi \!\simeq\! \partial_x\left(\phi^{m-1}\partial_x \phi \right) + \delta(x)f(t)\left(1-\phi\right) \ ,
\end{equation}
where we have used the fact that $\partial_x v\!\equiv\!\delta(x)f(t)$. Irrespective of the nonlinearity exponent $m$, the presence of the advection term $v\partial_x \phi$ suggests transforming the above into a reduced ODE [e.g., Eqs.~\eqref{eq:drop_1d_unentangled_ODE},~\eqref{eq:drop_1d_entangled_ODE}, ~\eqref{eq:2d_diffusion_unentangled} and ~\eqref{eq:2d_diffusion_entangled}] may involve translation in addition to rescaling. Additionally, combining the source together with the incompressibility $\phi_A + \phi_B\!=\!1$ renders the source's amplitude to depend on the local concentration $\phi$, as $\delta(x)f(t)\left(1-\phi\right)$.


With these features in mind, we consider again the dominant balance of the osmotic pressure with the concentration time-variation $\ell^2 \!\sim\!\phi^{m-1} t$, together with the increase in mass $\phi \ell \!\sim\!\int_0^t f(t')dt'$. To make progress, we assume $f$ has a temporal power-law variation $f\!=\!Q t^{q-1}$, allowing us to obtain the scaling relations for any power-law increase $q$, and to generalize for more complicated time-variations (potentially approximating those as piecewise power-law functions). Taking $Q t^{q-1}$ for our source, we obtain $\phi \ell \!\sim\! Q t^q$, from which we recover $\phi(x,t)\!\equiv\!Q^{\tilde{\gamma}_1}t^{\tilde{\delta}_1} \Phi(\xi)$ and $\xi\!\equiv\!x/Q^{\tilde{\gamma}_2}t^{\tilde{\delta}_2}$ with $p\!\equiv\!2+d(m-1)$ as defined above, and
\begin{equation}\label{eq:scaling_source}
\begin{aligned}
    &\tilde{\gamma}_1\!=\!2/p \ ,& \, \tilde{\delta}_1\!=\!(2q-d)/p \ , &\\
    & \tilde{\gamma}_2\!=\!(m-1)/p \ ,&  \, \tilde{\delta}_2\!=\!\left[1+q(m-1)\right]/p \ .& 
\end{aligned}
\end{equation}

Comparing Eq.~\eqref{eq:scaling_source} with the scaling relations in the passive droplet diffusion case, Eq.~\eqref{eq:scaling_diffusion}, reveals that only the time dependence, via the $\tilde{\delta}$ exponents, are modified by the presence of a source, in which case $Q$ plays a role reminiscent of $\mcI$ in the passive case. These scaling exponents are the ones shown in the late-time regimes in Fig.~\ref{fig:fig5}(b) and (d), signifying the droplet's expansion before saturation.


Combining the scaling and taking into account the advection term, we propose to use the 1D transformation $\phi\!=\!Q^{\tilde{\gamma}_1}t^{\tilde{\delta}_1} \Phi(\eta)$ and $\eta\!=\!\left[x - \frac{1}{2 q}Q t^{q}\right]/ Q^{\tilde{\gamma}_2}t^{\tilde{\delta}_2}\!\equiv\!\xi+\xi_s$, where we defined $\xi_s\!\equiv\!-\frac{Q^{1-\tilde{\gamma}_2} t^{q-\tilde{\delta}_2}}{2q}$ as the rescaled source coordinates (see App.~\ref{ap:source-driven_self-similarity}). Using these in Eq.~\eqref{eq:source_1d_simplified} (setting aside the source term, which we treat as a boundary condition to our equations), we obtain
\begin{gather}
    \left[3\Phi_u\Phi_u'+(1-2q)\eta\Phi_u\right]'+3 q \eta \Phi_u' = 0 \quad \text{for} \ \alpha\!=\!1 \ , \label{eq:ss_source_unentangled} \\
    \left[4\Phi_e^2\Phi_e'+(1-2q)\eta\Phi_e\right]' + 4q\eta\Phi_e' = 0 \quad   \text{for} \ \alpha\!\gg\!1 \ . \label{eq:ss_source_entangled} 
\end{gather}
Clearly, in both cases the emerging equations are cast as ordinary differential equations, however in the self-similar coordinates $\eta$ the source is moving back to $\xi_s$, and the boundary conditions themselves are time-dependent. So, while the equations are ordinary differential equations, the domain of integration is time dependent. This explicit time dependence breaks the self-similar structure.


\begin{figure}[t]
\includegraphics[width=0.48\textwidth]{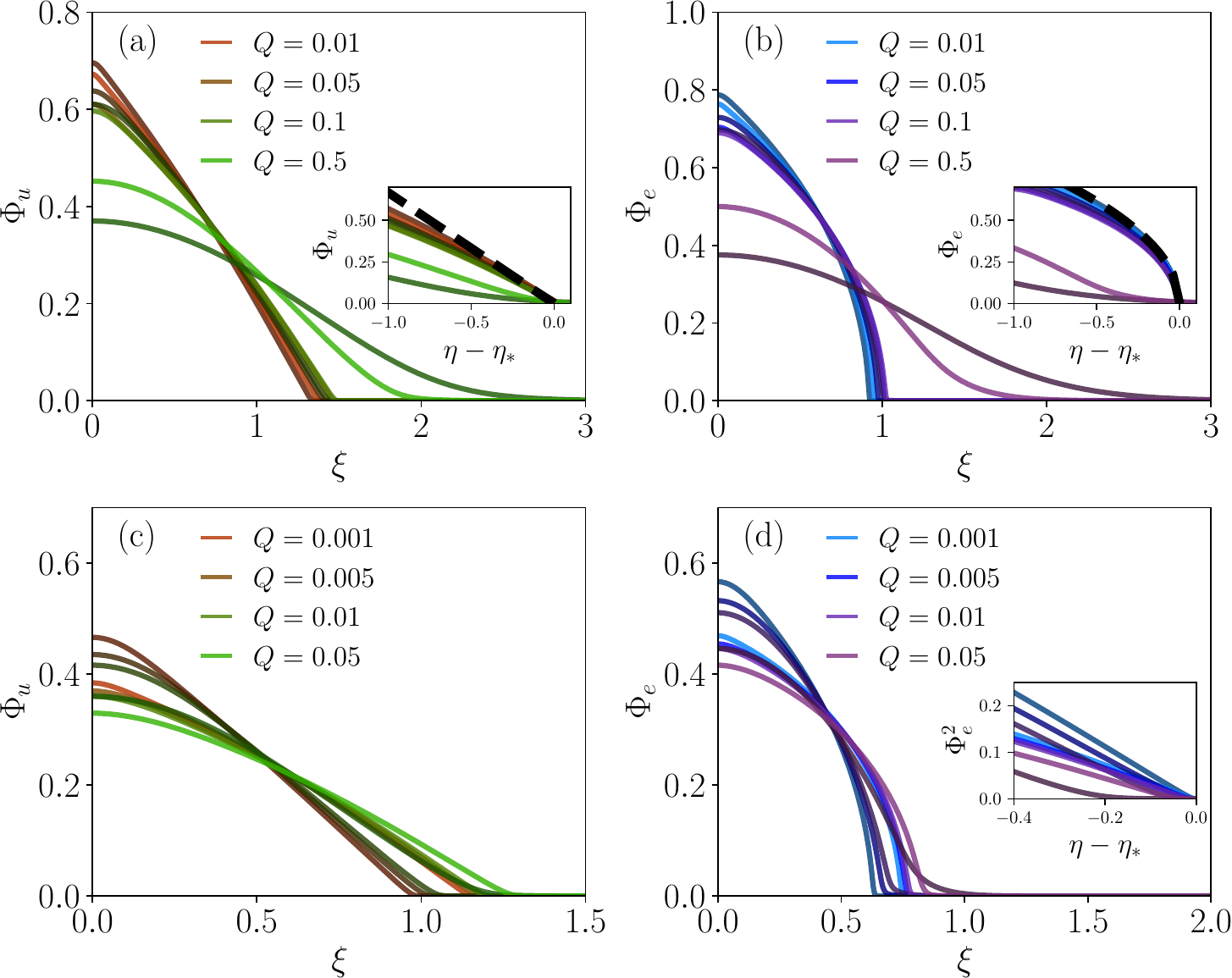}
\caption{Numerical solutions and analytical approximation for droplet fronts of the unentangled and entangled source-driven polymer interdiffusion under different time-dependent sources. (a) The rescaled solutions $\Phi_u\!=\!\phi / \left(Q^2t\right)^{1/3}$ in the rescaled coordinates $\xi\!=\!x/\left(Q t^2\right)^{1/3}$ for the unentangled case with time-independent source $q\!=\!1$, and for different amplitudes $Q$. Each different color shows profiles obtained for different $Q$ value, where for each $Q$ value, we show two different time instances in darker and lighter curves (the dark curves represent earlier times, and light curves later times). Inset: the same solutions shifted to their front coordinate $\eta-\eta_*$, with the theoretical prediction of $\Phi_u\!\sim\!-\frac{2}{3} \eta_* \eta$ (with $\eta_*\!\simeq\!1$ obtained from our numerical simulations). 
(b) The rescaled solution $\Phi_e\!=\!\phi / \left(Q^2 t\right)^{1/4}$ in the coordinates $\xi\!=\!x/\left(Q^2 t^3\right)^{1/4}$ for the entangled source-driven spreading, with $q\!=\!1$ for different source amplitudes $Q$. For each $Q$ value, we show two different time instances in darker and lighter curves (the dark curves represent earlier times, and light curves later times). Inset: the same solutions shifted to their front coordinate $\eta-\eta_*$, with the theoretical prediction of $\Phi_e\!\sim\!\sqrt{-\frac{3}{2} \eta_* \eta}$ (with $\eta_*\!\simeq\!0.5$ obtained from our numerical simulations).
(c) The rescaled solutions $\Phi_u\!=\!\phi / \left(Q^{2/3}t\right)$ in  $\xi\!=\!x/\left(Q^{1/3}t\right)$ coordinates for the unentangled case, under a time-dependent $q\!=\!2$ source-driven A-mer dynamics, as  obtained from our 1D numerical simulations. For each $Q$ value, we show two different time instances in darker and lighter curves (the dark curves represent earlier times, and light curves later times). (d) The rescaled solution $\Phi_e\!=\!\phi / \left(Q^2 t^3\right)^{1/4}$ in the coordinates $\xi\!=\!x  / \left(Q^2t^5\right)^{1/4}$ for entangled $\alpha\!=\!10^4$ A-mers for $q\!=\!2$. For each $Q$ value, we show two different time instances in darker and lighter curves (the dark curves represent earlier times, and light curves later times). Inset: $\Phi_e^2$ versus the front coordinates $\eta-\eta_*$ showing $\Phi_e\!\sim\!\sqrt{\eta_*-\eta}$.}
\label{fig:fig6}
\end{figure}

Instead of solving the ODEs of Eq.~\eqref{eq:ss_source_unentangled}-\eqref{eq:ss_source_entangled} with a time-dependent domain and boundary conditions, we perform an asymptotic expansion near the front $\eta_*$, where $\Phi(\eta_*)\!=\!0$. We anticipate that while the solutions' features near the source may vary, the droplets' fronts will attain a similar diffusion-dominated, spatial dependence as their passive counterparts, as already demonstrated in Fig.~\ref{fig:fig5}(a) and (c). Substituting the expansion $\tilde{\Phi}\simeq a_1 y^{1/(m-1)} + a_2 y^{2/(m-1)}$ (where $y\!\equiv\!\eta_*-\eta$)  in Eq.~\eqref{eq:ss_source_unentangled}-\eqref{eq:ss_source_entangled}, using $m\!=\!2$ and $m\!=\!3$ for $\Phi_u$ and $\Phi_e$ respectively, collecting the leading orders, and demanding that those vanish, lead to $a_1\!=\!\frac{2}{3}\eta_*$, and $a_1\!=\!\sqrt{\frac{3}{2}\eta_*}$ for $m\!=\!2$ and $m\!=\!3$ respectively.

We rescale the numerical solutions obtained for the unentangled and entangled cases, at different times and for different amplitudes $Q$, according to Eq.~\eqref{eq:scaling_source}, as shown in Fig.~\ref{fig:fig6}(a) and (b). For high $Q$ values, our numerical integration did not advance far enough to transition into the long-time scaling regime, but the other solutions converge onto similar curves. The obtained curves vary especially near the origin, but as the front is approached, the different solutions converge onto a similar profile. In the insets in Fig.~\ref{fig:fig6}(a) and (b), we show the rescaled solutions in $\eta$ coordinates, shifted to the front $\eta_*$, and plot the leading order expected behavior discussed above. Clearly, the well-developed fronts all agree with our predicted leading-order behavior.

The constant source with $q\!=\!1$ is a special case in which the source $\xi_s$ moves with the same time-dependence as the rescaled coordinates $\xi$. We now show that similar spatial profiles arise for $q\!=\!2$, i.e., with a time-dependent source. We plot the rescaled numerical solutions at various times and for different $Q$ values in Fig.~\ref{fig:fig6}(c) and (d), showing an explicit linear front geometry for the unentangled case in panel (c). In panel (d) we include an inset plotting $\Phi_e^2$ versus $\eta-\eta_*$, showing that $\Phi_e\!\sim\!\sqrt{\eta_*-\eta}$, similar to the $q\!=\!1$ case above.

\begin{figure}[t]
\includegraphics[width=0.48\textwidth]{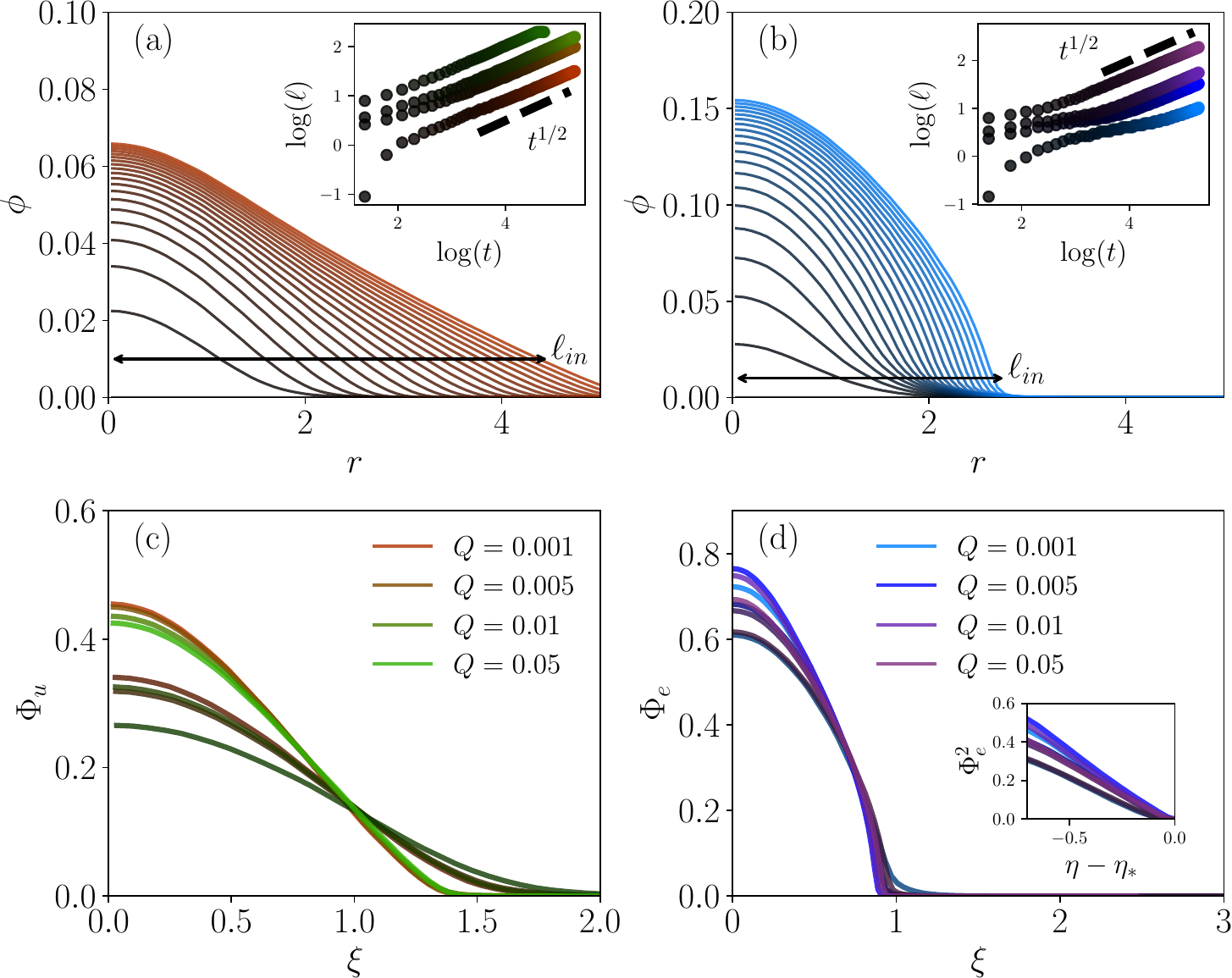}
\caption{Numerical solutions for unentangled and entangled source-driven polymer interdiffusion in 2D. (a) Radially-averaged $\phi$ profiles for 2D unentangled source-driven A-mers obtained from the numerical solution of Eqs.~\eqref{eq:con_clean}-\eqref{eq:rel_clean} ($\alpha\!=\!1$, $N_A\!=\!10^4$ and $Q\!=\!0.01$, early times are depicted as gray, and later times are drawn in red). Inset: the typical droplet radius as a function of time for different $Q$ values [see panel (c) for legends], with a power-law of $t^{1/2}$ as a guide to the eye. (b) Radially-averaged $\phi$ profiles for 2D entangled A-mers ($\alpha\!=\!10^4$, $N_A\!=\!10^4$ and $Q\!=\!0.01$, early times are depicted as gray, and later times are drawn in blue). Inset: the typical droplet radius as a function of time for different $Q$ values [see panel (d) for legends], with a power-law of $t^{1/2}$ as a guide to the eye.
(c) The rescaled solution $\Phi_u\!=\!\phi / Q^{1/2}$ in the rescaled coordinates $\xi\!=\!r/\left(Q t^2\right)^{1/4}$ (darker and lighter curves representing different times of the same trajectory). (d) The rescaled solution $\Phi_e\!=\!\phi / Q^{1/3}$, and $\xi\!=\!r/ \left(Q^2 t^3\right)^{1/6}$. Inset: $\Phi^2$ versus $\eta-\eta_*$ showing $\Phi_e\!\sim\!\sqrt{\eta_*-\eta}$.}
\label{fig:fig7}
\end{figure}

Finally, we perform 2D numerical simulations with $q\!=\!1$, and compare our results in a similar fashion in Fig.~\ref{fig:fig7}. Panels (a) and (b) show exemplary radially-averaged $\phi$ dynamics. In 2D, taking $q\!=\!1$ reveals an interesting case in which the source amplitude is time-independent [see $\tilde{\delta}_1$ in Eq.~\eqref{eq:scaling_source}], and $\tilde{\delta}_2\!=\!1/2$ independently of $m$, implying the unentangled and entangled time-dependencies are similar. We show this scaling in the insets of panels (a) and (b) respectively (here probing only a single length-scale $\ell_{in}$, as the half-width of the $\phi\!>\!0.01$ domain). Then, panels (c) and (d) (and the inset therein) show the rescaled solutions, revealing similar spatial features at the droplets' front, as the 1D counterparts (see Fig.~\ref{fig:fig6}).



\section{Discussion}\label{se:discussion}
Inspired by biocondensate diffusive and mixing dynamics, we used a two-fluid formalism to analyze polymer-polymer interdiffusion problems. We presented the equations of motion and demonstrated that under certain approximations the interdiffusion of a polymer droplet could be cast as a nonlinear diffusion equation, similar to diffusion in porous media~\cite{barenblatt2003scaling,zheng2022influence}. We highlighted the fact that the entanglement parameter $\alpha$ can dramatically alter the nonlinearity exponent $m$, changing the anticipated scaling and spatial structure of the solutions. The results obtained from our full numerical simulations in both 1D and 2D corresponded well with the anticipated self-similar analytical solutions for the two $\alpha$ limits considered for unentangled case $m\!=\!2$ (corresponding to $\alpha\!=\!1$), and the entangled case $m\!=\!3$ (corresponding to $\alpha\!\gg\!1$).

Then, we introduced a source into the equations of motion --- akin to a continuous synthesis of polymers, prevalent in biological contexts. We demonstrated that for this case there are multiple temporal scaling laws that characterize the spread at different time regimes, and that while the source alters the time-dependence of the solutions and breaks the self-similar structure, the local geometry near the front bears similarity to the ``passive diffusion'' cases. Using both 1D and 2D numerical solutions, we demonstrated this similarity persists for various $q$ values, and in different spatial dimensions $d$.

We suspect that the local front geometry could affect polymer conformations~\cite{ShortPaper}. The sharp interface of the droplets hints that the velocity profile should vary dramatically there, and should converge to a constant sufficiently far away. This local variation at the rim induces strain-rate gradients, and induce compressive forces on the polymers, causing them to adopt folded conformations. We suspect such effect could partially explain the compression of ribosomal RNA conformations as they drift away from the nucleolus~\cite{ShortPaper,riback2023viscoelasticity}.

In the above work we neglected elastic and viscous effects that could modify the shape of the formed droplets, and cause the solutions to deviate from the reduced self-similar solutions. Such coupling between osmotic pressure and mechanics lies at the heart of gel physics~\cite{tanaka1993unusual,tanaka1996universality,tanaka1997phase,tanaka1997viscoelastic,tanaka2000viscoelastic,tanaka2006viscoelastic,tanaka2012viscoelastic,tanaka2022viscoelastic}. While the source-driven polymer interdiffusion problem considered above is reminiscent of a driven Cahn–Hilliard system~\cite{cahn1958free,cahn1959free}, the presence of material stresses may introduce additional temporal dependencies and spatial patterns, especially near the center and edges of the expanding droplets.

It may be interesting to consider a source-driven problem for phase separating mixtures. In such cases, the osmotic pressure would oppose the outward flux, which would eventually lead to droplet formation. We suspect this scenario could also be relevant for biological phase separation~\cite{alberti2017phase,mitrea2016phase}. Overall, we believe the presented two-fluid framework, the analysis above, and future works would deepen our understanding of polymer interdiffusion physics and biophysical diffusion processes.

\emph{Acknowledgments. --- } A.M. acknowledges support from the \href{https://doi.org/10.37717/2021-3362}{James S. McDonnell Foundation Postdoctoral Fellowship Award in Complex Systems}. We acknowledge support from the NSF Grant DMS/NIGMS 2245850 and from the Princeton Center for Complex Materials (PCCM), an NSF-supported Materials Research Science and Engineering Center under award DMR-2011750.
The authors also thank N. Wingreen and C. P. Brangwynne for helpful comments and discussions.

\appendix
\section{Two-dimensional passive droplet diffusion}\label{ap:2D_formalism}
In 2D we assume that the flow is radial, implying the radial velocities $u^r_A$ and $u^r_B$ obey $\nabla\cdot\left(\phi u_A^{r} + \left(1-\phi\right)u_B^r\right)\equiv\!\nabla\cdot \bm{v}\!=\!0$, implying $v^r\sim \frac{v_0}{r}$. From here we use similar simplifications as those used for the 1D case, neglecting the material stresses $\bm{\sigma}_A$ and $\bm{\sigma}_B$, so that $w^r\equiv u_A^r - u_B^r\!\simeq\!-\frac{\left(1-\phi\right )}{\mu}\partial_r \Pi$ ($\bm{\Pi}\!\equiv\! \bm{I} \Pi$, assumed to be isotropic). Expressing $u_A^r\!=\!v^r + \frac{(1-\phi)^2}{\mu}\partial_r \Pi$, and using this back in Eq.~\eqref{eq:con_clean} (without a source) yields
\begin{equation}
    \partial_t \phi\!\simeq\! \frac{1}{r}\partial_r \left(r\phi^{m-1}\partial_r \phi\right)-v^r\partial_r\phi \ ,
\end{equation}
where we used $\nabla\cdot \bm{v}\!=\!0$. While $v^r\simeq \frac{v_0}{r}$ could contribute significantly near the origin, as we are examining a droplet expansion, we assume that at sufficiently late times when the droplet has extended, this term can be neglected, yielding a similar nonlinear diffusion equation as in the 1D case.

From here we proceed with the usual scaling arguments. The nonlinear equation implies $\ell^2\sim \phi^{m-1} t$, and the integral conservation law implies $\phi \ell^2\!\sim\mathcal{I}$, yielding $\phi\sim\left(\mcI/t\right)^{1/m}$, and $\ell\sim \mcI^{\frac{m-1}{2m}}t^{\frac{1}{2m}}$, as obtained from Eq.~\eqref{eq:scaling_diffusion} (with $d\!=\!2$).

\section{Higher-dimensional problems with a source}\label{ap:higherDs}
In higher spatial dimensions, we add Eq.~\eqref{eq:con_clean} to obtain
\begin{equation}\label{eq:incomp_higherD}
    \nabla\cdot\bm{v}\!=\!S \ .
\end{equation}
Now, using $\bm{w}$ and $\bm{v}$ to express $\bm{u}_A\!=\!\bm{v}+(1-\phi)\bm{w}$, we can rewrite 
\begin{equation}
    \partial_t \phi = \nabla \cdot \left[\frac{(1-\phi)^2\phi}{\mu}\nabla\cdot\bm{\Pi} - \bm{v} \phi  \right] + S \ ,
\end{equation}
analogous to Eq.~\eqref{eq:source_1d},
where here we again neglect the material stresses $\bm{\sigma}_A$ and $\bm{\sigma}_B$. We can then express this also as
\begin{equation}
    \partial_t \phi + \bm{v}\cdot\nabla\phi = \nabla \cdot \left[\frac{(1-\phi)^2\phi}{\mu}\nabla\cdot\bm{\Pi}\right] + S(1-\phi) \ ,
\end{equation}
analogous to Eq.~\eqref{eq:source_1d_simplified},
where we used Eq.~\eqref{eq:incomp_higherD}. This explicitly shows that the addition of a source implies an advection term, and a modification to the source amplitude, as was shown in 1D above.

\section{Transformation into self-similar variables for the source-driven problem}\label{ap:source-driven_self-similarity}

Here we provide detailed calculations that show how using the transformations $\phi\!=\!Q^{\tilde{\gamma}_1}t^{\tilde{\delta}_1} \Phi(\eta)$ and $\eta\!=\!\left[x - \frac{1}{2 q}Q t^{q}\right]/ Q^{\tilde{\gamma}_2}t^{\tilde{\delta}_2}$ proposed above in Eq.~\eqref{eq:source_1d} result in ordinary differential equation.

To begin we evaluate 
\begin{equation}
    \begin{gathered}
        \partial_t \eta = -\frac{1}{2} Q^{1-\tilde{\gamma}_2} t^{q-1-\tilde{\delta}_2} - \tilde{\delta}_2 t^{-1} \eta\ , \\
        \partial_x \eta = Q^{-\tilde{\gamma}_2}t^{-\tilde{\delta}_2} \ . 
    \end{gathered}
\end{equation}
Taking similar derivatives for $\phi$, we have
\begin{equation}
    \begin{gathered}
    \partial_t \phi = \tilde{\delta}_1 Q^{\tilde{\gamma}_1}t^{\tilde{\delta}_1-1} \Phi + Q^{\tilde{\gamma}_1}t^{\tilde{\delta}_1} \Phi' \partial_t \eta \ , \\ 
    \partial_x \phi =Q^{\tilde{\gamma}_1}t^{\tilde{\delta}_1} \Phi' \partial_x \eta .
    \end{gathered}
\end{equation}

Using the above relations in the left-hand side of Eq.~\eqref{eq:source_1d} (focusing on $x\!>\!0$ for simplicity), results in
\begin{equation}
    \partial_t \phi + \frac{1}{2} Qt^{q-1}\partial_x \phi = Q^{\tilde{\gamma}_1}t^{\tilde{\delta}_1-1}\left(\tilde{\delta}_1\Phi - \tilde{\delta}_2\eta \Phi'\right) \ .
\end{equation}
Similarly, using the above in the nonlinear diffusion term  yields
\begin{equation}
    \partial_x\left(\phi^{m-1}\partial_x \phi\right) = Q^{m\tilde{\gamma}_{1}-2\tilde{\gamma}_{2}}t^{m\tilde{\delta}_{1}-2\tilde{\delta}_{2}}\left[\left(m-1\right)\Phi'^2 + \Phi \Phi''\right]\Phi^{m-2} \ .
\end{equation}
Comparing the powers of these two expression reveals
\begin{equation}
    \begin{gathered}
       \left(m-1\right)\tilde{\gamma}_1 \!=\! 2\tilde{\gamma}_{2}\ , \\
       \left(m-1\right)\tilde{\delta}_1 \!=\!2\tilde{\delta}_{2}-1 \ ,
    \end{gathered}
\end{equation}
both obeyed by Eq.~\eqref{eq:scaling_source}. Using $m\!=\!2$ and $m\!=\!3$ yield Eqs.~\eqref{eq:ss_source_unentangled}-\eqref{eq:ss_source_entangled} respectively.

\bibliography{polymers.bib} 
\end{document}